\magnification \magstep1
\input amssym.def
\input amssym.tex
\def \sf{\Sigma_F}
\def \si{\Sigma_I}
\def \ep{\epsilon}

\def \rta{\rightarrow}
\def \ms{\medskip}

\def \ss{\smallskip}
\def \sclass{S_{\rm class}}

\def \alf{\alpha}

\medskip
\centerline{\bf Coherent and Squeezed States in Black-Hole Evaporation}
\bigskip
\centerline{ A.N.St J.Farley and P.D.D'Eath }
\medskip
\centerline{Department of Applied Mathematics and Theoretical Physics, 
Centre for Mathematical Sciences,}
\smallskip
\centerline{University of Cambridge, Wilberforce Road, 
Cambridge CB3 0WA, United Kingdom}
\medskip
\centerline{Abstract} 
\medskip 
\noindent 
In earlier Letters, we adopted a complex approach 
to quantum processes in the formation and evaporation 
of black holes.  Taking Feynman's $+{\,}i\ep$ prescription, 
rather than one of the more usual approaches, 
we calculated the quantum amplitude 
(not just the probability density) for final weak-field 
configurations following gravitational collapse 
to a black hole with subsequent evaporation.  
What we have done is to find quantum amplitudes 
relating to a pure state at late times following 
black-hole matter collapse.  Such pure states 
are then shown to be susceptible to a description 
in terms of coherent and squeezed states 
-- in practice, this description is not very different 
from that for the well-known highly-squeezed final state 
of the relic radiation background in inflationary cosmology.  
The simplest such collapse model involves Einstein gravity 
with a massless scalar field.  The Feynman approach 
involves making the boundary-value problem for gravity 
and a massless scalar field well-posed.  
To define this, let $T$ be the proper-time separation, 
measured at spatial infinity, between two space-like 
hypersurfaces on which initial (collapse) 
and final (evaporation) data are posed.  
Then, in this approach, one rotates 
$T\to{\mid}T{\mid}\exp(-{\,}i\delta)$
into the lower half-plane.  
In an adiabatic approximation, the resulting quantum 
amplitude may be expressed in terms of generalised 
coherent states of the quantum oscillator, 
and a physical interpretation is given.  
A squeezed-state representation, as above, 
then follows.\par
\medskip
\noindent 
{\bf 1. Introduction}
\medskip
\indent 
We begin by describing Feynman's 
$+{\,}i\ep$ approach [1] in the context 
of black-hole evaporation. 
In [2-12], this treatment was described 
and applied to the calculation of quantum amplitudes 
(not just probabilities) for particle production, 
following gravitational collapse to a black hole.  
Suppose, for definiteness, that one's Lagrangian 
contains Einstein gravity coupled to a real massless 
scalar field.  Asymptotically-flat initial data 
are posed on an initial space-like 
hypersurface $\si{\,}$, and final data 
on a surface $\sf{\,}$, separated from $\si$ 
by a (large) real Lorentzian time-interval $T{\,}$, 
as measured at spatial infinity.  
Suppose further, for simplicity, 
that the initial data on $\si$ 
are spherically symmetric, 
corresponding to a diffuse slowly-moving 
initial matter distribution.  
The final data for gravity + scalar are taken 
to have a `background' spherically-symmetric part, 
plus small non-spherical perturbations, 
which correspond to gravitons 
and massless-scalar particles.\par 
\ss 
\indent 
Following Feynman's $+{\,}i\epsilon$ procedure [1], 
one rotates the time-interval $T$ into the complex:
${\,}T\rta{\mid}T{\mid}\exp(-{\,}i\delta){\,}$, with 
$0<\delta\leq\pi/2{\,}$.  
The classical boundary-value problem, 
for a complex 4-metric $g_{\mu\nu}$ 
and scalar field $\phi$ given the above data 
on $\si{\,},\sf{\,}$, is then expected 
to be well-posed, unlike the ill-posed case 
${\,}\delta =0{\,}$ 
(or equivalently $T$ real) [3,13,14].  
One can evaluate the second-variation classical 
action $S^{(2)}_{\rm class}$ as a functional 
of the (still real) boundary data and as a function 
of the complex variable $T{\,}$.  
One then computes the corresponding semi-classical 
quantum amplitude, proportional to 
$\exp\bigl(iS^{(2)}_{\rm class}\bigr){\,}$, 
and can also include loop corrections, 
if appropriate.  
Finally, the Lorentzian quantum amplitude 
for black-hole evaporation 
(again, not just the probability density) 
is recovered by taking the limit 
as ${\,}\delta\rta 0_{+}{\,}$.\par 
\ss 
\indent 
In this Letter, we study such black-hole evaporation 
amplitudes, which were constructed in detail 
in [5-8, 10-12], but now in the context 
of coherent states [15], 
which resemble `classical states', 
and of squeezed states [16], which are purely 
quantum-mechanical.  Although our motivation originated 
with the question of black-hole radiation, 
there are also strong connections between this work 
and the study of the relic Cosmic Microwave Background 
Radiation (CMBR) induced by inflationary cosmological 
perturbations.\par 
\ss 
\indent 
In inflationary cosmology, the field modes 
are in their adiabatic ground state, 
with short wavelengths near the start of inflation.  
Due to the accelerated expansion of the universe 
during inflation, quantum fluctuations are amplified 
into macroscopic or classical perturbations.  
The early-time fluctuations lead to the formation 
of large-scale structure in the universe, 
and also contribute to the anisotropies in the CMBR.  
The final state for the perturbations is a two-mode 
highly-squeezed state for modes whose radius 
is much greater than the Hubble radius [17], 
pairs of field quanta being produced at late times 
with opposite momenta.  Tensor $(s=2)$ fluctuations 
in the metric, for example, are predicted 
to give rise to relic gravitational waves.  
By comparison, electromagnetic waves $(s=1)$ 
cannot be squeezed in the same way.\par 
\ss 
\indent 
In either case, cosmological or black-hole, 
one works within an adiabatic approximation 
for the perturbative modes.  
Writing $k$ for a typical perturbative frequency, 
one requires $k{\,}{\gg}{\,}H$ in the cosmological case, 
where $H=(\dot a/a)$ and $a(t)$ is the scale factor.  
In the black-hole case, the space-time geometry 
at late times, in the region containing a stream 
of outgoing radiation, 
is given by a Vaidya metric [4,8,18,19] 
with a slowly-varying `mass function' $m(t,r){\,}$.  
The adiabatic condition then reads 
${\,}k{\,}{\gg}{\,}{\mid}{\dot m}/m{\mid}{\,}$.\par 
\ss 
\indent 
In applying the squeezed-state formalism, 
one finds, in the case of cosmological perturbations, 
that these evolve essentially according to a set 
of Schr\"odinger equations [20].  
Such perturbations, whether of density, 
rotational or gravitational type, 
starting in an initial vacuum state, 
are transformed into a highly-squeezed vacuum state, 
with many particles, having a large variance 
in their amplitude (particle number), 
but small (squeezed) phase variations.  
The squeezing of cosmological perturbations 
may be suppressed at small wavelengths, 
but it should be present at long wavelengths, 
especially for gravitational waves [21].  
These perturbations also induce the anisotropies 
at large angular scales, as observed in the CMBR.  
Their wavelengths today are comparable with or greater 
than the Hubble radius.  The above amplification 
of the initial zero-point fluctuations gives rise 
to standing waves with a fixed phase, 
rather than travelling waves.  
The relic perturbations in the high-squeezing 
or WKB limit can be described as a stochastic 
collection of standing waves.  
Although this paragraph has reviewed 
the application to cosmology, a similar picture 
emerges in the application 
to black-hole evaporation.\par 
\ss 
\indent 
Sec.2 outlines the main features of the above complex 
approach to the calculation of quantum amplitudes 
(not just probabilities) for data 
(spins $s=0{\,},1{\,},2$) prescribed on a late-time 
final hypersurface $\Sigma_{F}{\,}$.  
This requires a rotation: 
$T\rightarrow{\mid}T{\mid}\exp(-{\,}i\delta)$ 
into the lower half-plane.  The resulting amplitudes 
are then related to coherent and squeezed states.  
Secs.3,4,5 describe coherent states, 
generalised coherent states and squeezed states, 
respectively.  In Sec.6, the small angle $\delta$ 
(above), through which the time $T$ at infinity 
is rotated into the complex, 
is related to the large amount of squeezing 
which has been applied to give the final state.  
Sec.7 contains a brief Conclusion.\par 
\ms 
\noindent 
{\bf 2. The quantum amplitude for late-time data} 
\ms 
\indent 
Consider first the case of a rotation into the complex 
of the time-interval $T{\,}$, measured at spatial infinity, 
by a moderately small angle $\delta{\,}$, as above.  
One expects that the resulting classical solution 
$(g_{\mu\nu}{\,},\phi)$ of the coupled 
Einstein/massless-scalar field equations is slightly 
complexified, by comparison with a Lorentzian-signature 
solution.  By suitable choice of coordinates 
$(t,r,\theta,\varphi)$, the spherically-symmetric 
'background' part of the metric may be written 
in the form [2,6]
$$ds^{2}{\;}{\,}
={\;}-{\,}e^{b}{\,}dt^{2}{\,}+{\,}e^{a}{\,}dr^{2}{\,}
+{\,}r^{2}{\,}\bigl(d{\theta}^{2}
+{\,}{\sin}^{2}{\theta}{\,}d{\varphi}^{2}\bigr)
{\quad},\eqno(2.1)$$
where ${\,}b={\,}b(t{\,},r){\;},{\;}a={\,}a(t{\,},r){\,}$, 
and the spherically-symmetric 'background' part $\Phi$ 
of the scalar field has the form $\Phi =\Phi(t{\,},r){\,}$.  
The coupled Lorentzian-signature Einstein/scalar 
field equations for this spherically-symmetric 
configuration are given by the analytic continuation 
of the Riemannian field equations Eqs.(3.7-11) of [5], 
on making the replacement 
$$t{\;}{\,}
={\;}{\,}\tau{\,}\exp(-{\,}i{\,}\vartheta)
{\quad},\eqno(2.2)$$ 
where $\tau$ is the 'Riemannian time-coordinate' of [5], 
and where the real number $\vartheta$ should be rotated 
from $0$ to ${\pi}/2{\,}$.\par  
\smallskip    
\indent
Small non-spherical perturbations in the boundary data 
given on the final late-time hypersurface ${\Sigma}_{F}$
consist of the perturbed part of the intrinsic 3-dimensional 
spatial metric $h_{ijF}$ on $\Sigma_{F}{\,}$, together 
with the perturbations in the scalar field ${\phi}_{F}$
on $\Sigma_{F}{\,}$.  As above, these correspond to gravitons 
and to massless-scalar particles, propagating 
on the spherically-symmetric classical background 
$\bigl(g_{\mu\nu}{\,},\Phi\bigr){\,}$.  For example, 
the linearised scalar perturbations $\phi^{(1)}{\,}$,
given [2] by ${\,}\phi{\,}={\,}\Phi +\phi^{(1)}{\,}$, 
may be first decomposed as in Eq.(6) of [2], namely as:
$$\phi^{(1)}(t{\,},r{\,},\theta{\,},\varphi){\;}{\,}
={\;}{\,}{{1}\over{r}}{\;}
\sum_{\ell =0}^{\infty}{\,}\sum_{m=-\ell}^{m=\ell}{\;}
Y_{\ell m}(\Omega){\;}R_{\ell m}(t{\,},r)
{\quad}.\eqno(2.3)$$
Here, $Y_{\ell m}(\Omega)$ denotes the $(\ell{\,},m)$ 
spherical harmonic of [22].  The scalar field equation
decouples for each $(\ell{\,},m)$, leading to the mode equation
$$\Bigl(e^{(b-a)/2}{\,}{\partial}_{r}\Bigr)^{2}
R_{\ell m}{\,}
-{\,}\Bigl(\partial_{t}\Bigr)^{2}R_{\ell m}{\,}      
-{\,}{{1}\over{2}}{\,}\Bigl(\partial_{t}\bigl(a-b\bigr)\Bigr){\,}
\Bigl(\partial_{t}R_{\ell m}\Bigr){\,}
-{\,}V_{\ell}(t,r){\,}R_{\ell m}{\;}{\,}
={\;}{\,}0
{\quad},\eqno(2.4)$$
where
$$V_{\ell}(t{\,},r){\;}{\,}
={\;}{\,}{{e^{b(t{\,},r)}}\over{r^{2}}}{\;}
\biggl(\ell(\ell +1){\,}+{\,}{{2m(t{\,},r)}\over{r}}\biggr)
\eqno(2.5)$$
is real and positive in the Lorentzian-signature case.
The 'mass function' $m(t{\,},r){\,}$, which would equal 
the constant mass $M$ for an exact Schwarzschild 
geometry [23], is defined by
$$e^{-{\,}a(t{\,},r)}{\;}{\,}
={\;}{\,}1{\,}-{\,}{{2m(t{\,},r)}\over{r}}
{\quad}.\eqno(2.6)$$
An analogous harmonic decomposition can be given 
for weak gravitational-wave perturbations 
about the spherical background [10].\par 
\smallskip
\indent
In most regions of the classical space-time, 
except for the central region where the black hole 
is formed, the metric functions $a(t{\,},r)$ and $b(t{\,},r)$ 
vary 'slowly' or 'adiabatically'.  In this case, 
one can consider a radial mode solution for (say) 
a perturbed scalar field, of the form [6]
$$R_{\ell m}(t{\,},r){\quad}
\sim{\quad}\exp(ikt){\;}{\;}\xi_{k\ell m}(t{\,},r)
{\quad},\eqno(2.7)$$
where ${\,}\xi_{k\ell m}(t{\,},r)$ varies 'slowly' with respect 
to $t{\,}$.  This will occur near spatial infinity, 
and it will also occur, provided that the time-interval $T$ 
is sufficiently large, in a neighbourhood of the final 
hypersurface $\Sigma_{F}{\,}$.  The mode equation (2.4,5) 
then reduces [6] to 
$$e^{(b-a)/2}{\;}{{\partial}\over{\partial r}}
\Biggl(e^{(b-a)/2}{\;}{{\partial\xi_{k\ell m}}\over{\partial r}}\Biggr)
{\;}+{\;}\Bigl(k^{2}{\,}-{\,}V_{\ell}\Bigr){\,}\xi_{k\ell m}{\;}{\,}    
={\;}{\,}0
{\quad}.\eqno(2.8)$$
The spherically-symmetric background metric in this region 
can be represented to high accuracy by a Vaidya metric [8,18,19], 
which describes the (on average) spherically-symmetric outflow 
of massless matter.  The principal condition for the validity 
of the adiabatic expansion is [6] that
$${\mid}k{\mid}{\quad}
{\gg}{\quad}{\mid}{\dot m}/m{\mid}
{\quad}.\eqno(2.9)$$   
\indent
In analysing the behaviour of the radial mode equation (2.8),
it is natural to define a generalisation $r^{*}$ of the standard
Regge-Wheeler or 'tortoise' coordinate ${\,}r_{S}^{*}{\,}$ 
for the Schwarzschild geometry [23], according to
$${{\partial}\over{\partial r^{*}}}{\;}{\,}
={\;}{\,}e^{(b-a)/2}{\;}{\,}{{\partial}\over{\partial r}}
{\quad}.\eqno(2.10)$$ 
The approximate (adiabatic) mode equation (2.8) then reads
$${{\partial^{2}\xi_{k\ell m}}\over{\partial r^{*2}}}{\;}
+{\;}\Bigl(k^{2}{\,}-{\,}V_{\ell}\Bigr){\,}\xi_{k\ell m}{\;}{\,}
={\;}{\,}0
{\quad}.\eqno(2.11)$$
\indent
We consider here, for definiteness, a set of suitable radial 
functions $\{\xi_{k\ell m}(r)\}$ on the final surface 
$\Sigma_{F}{\,}$, since it is here that the non-trivial 
boundary data are posed.  Since the mode equation (2.11) does not
depend on the quantum number $m{\,}$, we may choose 
$\xi_{k\ell m}(r){\,}={\,}\xi_{k\ell}(r){\;}$, independently
of $m{\,}$.  The boundary condition of regularity at the spatial 
origin $\{r=0\}{\;}$ [6] implies that 
$$\xi_{k\ell}(r){\;}{\,}
={\;}{\,}{\rm constant}{\,}{\times}{\,}\bigl(kr\bigr)^{\ell +1}{\;}
+{\;}O\Bigl(\bigl(kr\bigr)^{\ell +3}\Bigr)
\eqno(2.12)$$
as ${\,}r{\,}\rightarrow{\,}0_{+}{\,}$.  For the boundary 
condition on the ${\,}\xi_{k\ell}(r){\,}$ 
as ${\,}r\rightarrow\infty{\,}$, note that the potential 
${\,}V_{\ell}(r)$ decreases sufficiently rapidly, 
as $r\rightarrow\infty{\,}$, that a real solution 
to Eq.(2.11) behaves near $\{r=\infty\}$ according to
$$\xi_{k\ell}(r){\quad}
\sim{\quad}\biggl(z_{k\ell}{\;}\exp\Bigl(ikr_{S}^{*}\Bigr){\,}
+{\;}z_{k\ell}^{*}{\;}\exp\Bigl(-ikr_{S}^{*}\Bigr)\biggr)
{\quad}.\eqno(2.13)$$
Here, the $z_{k\ell}$ are certain dimensionless complex 
coefficients, which must be determined by using the differential
equation (2.11) together with the regularity conditions.  
Further [6], there is a natural normalisation of the basis 
$\{\xi_{k\ell}(r)\}$ of radial wave-functions.\par 
\smallskip
\indent
We continue, for purposes of exposition, to study the case 
of scalar perturbations, with a slightly complexified 
time-interval at infinity, $T={\mid}T{\mid}\exp(-{\,}i\delta){\,}$,
for $0<\delta{\;}{\leq}{\;}{\pi}/2{\,}$.  The relevant boundary 
data for anisotropic perturbations ${\,}\phi^{(1)}$ 
of the scalar field $\phi_{F}$ on $\Sigma_{F}{\,}$ 
can be described [6] by expanding out the interior classical
boundary-value solution near $\Sigma_{F}{\,}$ in the form
$$\phi^{(1)}{\;}{\,}
={\;}{\,}{{1}\over{r}}{\;}\sum_{\ell =0}^{\infty}{\;}
\sum_{m=-\ell}^{\ell}{\;}\int_{-\infty}^{\infty}{\;}
dk{\;}{\;}a_{k\ell m}{\;}{\,}\xi_{k\ell}(t{\,},r){\;}
{{\sin(kt)}\over{\sin(kT)}}{\;}Y_{\ell m}(\Omega)
{\quad}.\eqno(2.14)$$
Here, the real quantities $\{a_{k\ell m}\}$ characterise 
the final data.\par
\smallskip
\indent
More generally, for perturbative boundary data for a field 
of any spin, posed on $\Sigma_{F}{\,}$ in describing 
a final state resulting from black-hole evaporation, 
we denote by $\{a_{sk\ell mP}\}$ a set of analogous 
'Fourier-like' coefficients, where $s$ gives the particle 
spin, $k$ the frequency, $(\ell{\,},m)$ the angular quantum 
numbers, and $P={\pm}1$ the parity (for $s{\,}{\neq}{\,}0{\,}$).  
For massless perturbations of spins ${\,}s=0{\,},1{\,},2{\;}{\,}$ 
[2,3,5-7,10,11], we found that the quantum amplitude 
or wave functional is of semi-classical form, 
being given by
$$\Psi\Bigl[\{a_{sk\ell mP}\};{\,}T\Bigr]{\;}{\,}
={\;}{\,}N\,\exp\Bigl(i{\,}\sclass 
\Bigl[\{a_{sk\ell mP}\};{\,}T\Bigr]\Bigr)
{\quad},\eqno(2.15)$$  
where the pre-factor $N$ depends only on $T{\,}$.  Here,
$S_{\rm class}$ denotes the (second-variation) action 
of the classical infilling solution, as a functional 
of the boundary data.  For simplicity, we denote 
the collection $a_{sk\ell mP}{\,}$ of indices by $j{\,}$.  
Further, we write $M_{I}$ for the total (time-independent) 
ADM (Arnowitt-Deser-Misner) mass of the 'space-time', 
as measured at spatial infinity [23].  The ADM mass 
$M_{I}{\,}$, which is the limit at large radius 
of the variable mass $m(t{\,},r)$ of the Vaidya metric, 
is a functional of the final field configurations 
$\{a_{j}\}$ on $\Sigma_{F}{\,}$, since it depends on the full
gravitational field which results from classical solution 
of the complexified boundary-value problem.\par 
\smallskip
\indent 
As was found (for example) in the scalar case $s=0{\,}$ 
in [2,3,6], the classical action is dominated 
by contributions from frequencies $k$ with the values
$$k{\;}{\,}={\;}{\,}k_{n}{\;}{\,}
={\;}{\,}{{n\pi}\over{T}}{\quad};
{\qquad}{\qquad}{\qquad}n=1{\,},2{\,},3{\,},\ldots
{\quad}.\eqno(2.16)$$ 
We also define $\Delta k_{j}{\,}$ to be the spacing 
between neighbouring $k_{j}$--values:
$$\Delta k_{j}{\;}{\,}
={\;}{\,}{{\pi}\over{T}}
{\quad}.\eqno(2.17)$$
\indent
Following [2,3,5-7,10,11], the classical action functional 
$S_{\rm class}{\,}$ is found to be a sum over individual 
'harmonics' labelled by $j{\,}$, which depend 
on the corresponding indices $\{s{\,}k_{j}{\,}\ell{\,}m{\,}P\}$ 
through the quantity ${\mid}A_{j}{\mid}^{2}{\,}$, defined by
$${\mid}A_{j}{\mid}^{2}{\;}{\,}
={\;}{\,}2{\;}(-1)^{s}{\;}c_{s}{\;}{\,}
{{(\ell -s)!}\over{(\ell +s)!}}{\;}{\;}
{\mid}z_{j}{\mid}^{2}{\;}{\,}
\Bigl{\arrowvert}a_{j}{\,}+{\,}(-1)^{s}{\,}P{\,}a_{s,-k_{j}\ell m P}
\Bigl{\arrowvert}^{2}
{\quad}.\eqno(2.18)$$ 
Here, the coefficients $c_{s}{\,}$ for bosonic spins $s$ 
are given by ${\,}c_{0}=2\pi{\,},{\,}c_{1}={{1}\over{4}}{\,},
{\,}c_{2}={{1}\over{8}}{\,}$.  The quantities $z_{j}{\,}$ 
are the complex numbers appearing in Eq.(2.13), 
which arise in solving the adiabatic radial mode 
equation (2.11).  This leads to the form of the quantum amplitude:
$$\Psi\Bigl[\{A_{j}\};{\,}T\Bigr]{\;}{\,}
={\;}{\,}{\hat N}{\;}{\,}e^{-iM_{I}T}{\;}{\,}
\prod_{j}{\,}\Psi(A_{j}{\,};{\,}T)
{\;}{\,},\eqno(2.19)$$
where ${\hat N}$ also depends only on $T{\,}$.\par 
\smallskip
\indent
Taking the classical action $S_{\rm class}$ in the form found 
in [6] for the scalar $s=0$ case (for example), one deduces 
that the wave functional for given boundary data can be written:
$$\Psi\Bigl[\{A_{j}\}:{\,}T\Bigr]{\;}{\,}
={\;}{\,}{\hat N}{\;}{\,}e^{-iM_{I}T}{\;}{\,}
\prod_{j}{\;}{{1}\over{2i\sin(k_{j}T)}}{\;} 
\exp\biggl[{\,}{{i}\over{2}}{\,}
\bigl(\Delta k_{j}\bigr){\,}k_{j}{\,}
{\mid}A_{j}{\mid}^{2}{\,}\cot(k_{j}T)\biggr]
{\;}{\,}.\eqno(2.20)$$
This will be related to the coherent-state description
in the following Section 3.\par
\medskip
\noindent
{\bf 3. Coherent States}
\medskip
\indent
It is possible to rewrite the quantum amplitude (2.20)
with the help of the Laguerre polynomials [24].  
First, we introduce the associated Laguerre polynomials 
$L^{(m-n)}_{n}(x){\,}$, defined by
$$L^{(m-n)}_{n}(x){\;}{\,}
={\;}{\,}\sum^{n}_{p=0}{\;}{m\choose n-p}{\;}{{(-x)^{p}}\over{p!}} 
\eqno(3.1)$$ 
for $m\geq n\geq 0{\,}$.  The Laguerre polynomials $L_{n}(x)$ [24] 
are given by
$$L_{n}(x){\;}{\,}
={\;}{\,}L^{(0)}_{n}(x)
{\quad}.\eqno(3.2)$$ 
The set $\{L_{n}(x)\}$ obeys the completeness relation
$$\sum^{\infty}_{n=0}{\;}e^{-(x/2)}{\;}L_{n}(x){\;}
e^{-(y/2)}{\;}L_{n}(y){\;}{\,}
={\;}{\,}\delta(x,y)
{\quad}.\eqno(3.3)$$ 
Writing ${\,}z=x+iy{\,}$, consider now the function 
$L_{n}\bigl({\mid}z{\mid}^{2}\bigr){\,}$, 
which appears in Eq.(3.5) below.  For $n>0{\,}$, this cannot 
be written as a product of two (decoupled) wave functions 
of $x$ and $y$ in an excited state, due to pair 
correlations [25].  But, in terms of Hermite polynomials 
$H_{p}(x)$ [24], one can write
$$L_{n}(x^{2}+y^{2}){\;}{\,}
={\;}{\,}{{(-1)^{n}}\over{2^{2n}{\,}n!}}{\quad} 
\sum^{n}_{p=0}{\;}{n\choose p}{\;}H_{2p}(x){\;}H_{2n-2p}(y)
{\quad}.\eqno(3.4)$$ 
\indent 
From this, one can further decompose the quantum amplitude 
(2.20) as
$$\Psi\Bigl[\{A_{j}\};T\Bigr]{\,}
={\,}{\hat N}{\,}e^{-iM_{I}T}{\,} 
\exp\Bigl({-\Sigma_{j}\bigl(\Delta k_{j}\bigr)
k_{j}{\,}{\mid}A_{j}{\mid}^{2}/2}\Bigr) 
\prod_{j}\sum^{\infty}_{n=0}e^{-2iE_{n}T}{\,}L_{n}
\Bigl[k_{j}\bigl(\Delta k_{j}\bigr){\,}
{\mid}A_{j}{\mid}^{2}\Bigr],\eqno(3.5)$$  
where $E_{n}=\bigl(n+{{1}\over{2}}\bigr){\,}k_{j}{\,}$ 
is the quantum energy of the linear harmonic oscillator.  
Note also the dependence of the quantum amplitude 
on ${\mid}A_{j}{\mid}$ --- it is spherically symmetric.\par
\smallskip
\indent
The Schr\"odinger-picture wave functions
$$\Psi_{nj}(x_{j}{\,},T){\;}{\,}
={\;}{\,}{{N}\over{\pi}}{\;}e^{-(x_{j}/2)}{\;}{\,}
e^{-2iE_{n}T}{\;}{\,}L_{n}(x_{j})
{\quad}\eqno(3.6)$$  
appear in the wave-function (3.5), with
${\,}x_{j}=k_{j}{\,}(\Delta k_{j}){\;}{\mid}A_{j}{\mid}^{2}{\,}$. 
The wave functions (3.6) have a strong connection 
with the exact solution of the forced-harmonic-oscillator 
problem [26], with Hamiltonian
$$H{\;}{\,}
={\;}{\,}{{p^{2}}\over{2\mu}}{\,}
+{\,}{{1}\over{2}}{\,}\mu{\,}\omega^{2}q^{2}{\,}
+{\,}q{\,}F(t)
{\quad},\eqno(3.7)$$  
where $F(t)$ denotes an external force, $\mu$ the oscillator mass 
and $\omega $ the oscillator frequency.  Assume that $F(t)=0{\,}$ 
for $t<t_{0}$ and for $t>T{\,}$, so that the asymptotic states, 
at early and late times $t{\,}$, are free-oscillator states.  
One can calculate the amplitude $A_{km}$ to make a transition 
from the free-oscillator state ${\mid}m>$ (with $m$ particles) 
at early times $t<t_{0}{\,}$, to the free-oscillator state 
${\mid}k>$ at late times $t>T{\,}$.  Define the `Fourier transform' 
of the force:
$$\beta{\;}{\,}
={\;}{\,}\int^{T}_{t_0}dt{\;}F(t){\;}e^{-i\omega t}
{\quad},\eqno(3.8)$$ 
and set
$$z{\;}{\,}
={\;}{\,}{{{\mid}\beta{\mid}^{2}}\over{2{\,}\mu{\,}\omega}}
{\quad}.\eqno(3.9)$$  
It has been shown [27-29], in the case $m\geq k{\,}$, that
$$A_{km}{\;}{\,}
={\;}{\,}e^{i\lambda}{\;}e^{-(z/2)}{\;}
\biggl({{k!}\over{m!}}\biggr)^{{1}\over{2}}{\;} 
\biggl({{i\beta}\over{\sqrt{2{\,}\mu{\,}\omega}}}\biggr)^{m-k}{\;}
L_{k}^{(m-k)}(z)
{\quad},\eqno(3.10)$$  
where $\lambda$ is a real phase.  This expression 
also gives $A_{km}$ for $m\leq k{\,}$, 
since $A_{km}=A_{mk}$ is symmetric.\par
\smallskip
\indent 
In the adiabatic limit, in which the force $F(t)$ changes 
extremely slowly, one has $z{\,}{\ll}{\,}1{\,}$, 
and from general considerations a state which begins 
as ${\mid}k>$ must end up in the same state ${\mid}k>$ 
after the time-dependent force has been removed.  
From Eq.(3.10), one has
$$A_{kk}{\;}{\,}
={\;}{\,}e^{i\lambda}{\;}e^{-(z/2)}{\;}L_{k}(z)
{\quad}.\eqno(3.11)$$ 
The corresponding probability that there should 
be no change in the number of particles is
${\,}{\mid}A_{kk}{\mid}^{2}{\,}={\,}e^{-z}{\;}[L_{k}(z)]^{2}{\;}$. 
Apart from the introduction of mode labels $j$ denoting 
the `quantum numbers' $\{sk\ell mP\}$, together with a 
necessary re-interpretation for $z{\,}$, these amplitudes 
are effectively the wave functions (3.5) derived 
from our boundary-value problem.\par
\smallskip
\indent 
One further viewpoint can be brought to bear on Eq.(3.10), 
arising from the coherent-state representation.  
Coherent states ${\mid}\alf>$ can be regarded 
as displaced vacuum states; that is, [15]
$${\mid}\alpha>{\;}{\,}
={\;}{\,}D(\alpha){\,}{\mid}0>
{\quad},\eqno(3.12)$$ 
where
$$D(\alpha){\;}{\,}
={\;}{\,}\exp\Bigl(\alpha{\,}a^{\dagger}{\,}
-{\,}\alpha^{*}a\Bigr)
\eqno(3.13)$$  
is a unitary displacement operator, obeying
$$D^{\dagger}(\alpha){\;}={\;}D^{-1}(\alpha){\;}{\,}
={\;}{\,}D(-{\,}\alpha)
{\quad},\eqno(3.14)$$  
and where the states ${\mid}\alpha>$ are eigenstates 
of the annihilation operator $a$ with complex eigenvalue 
$\alpha{\,}$.  Among quantum states for the harmonic 
oscillator, they are the closest to classical states, 
in that they attain the minimum demanded 
by the uncertainty principle.   Coherent states 
form an over-complete set, and are not orthogonal.  
In terms of the Fock-number eigenstates
$${\mid}n>{\;}{\,}
={\;}{\,}{{(a^{\dagger})^{n}}\over{\sqrt{n!}}}{\;}{\,}{\mid}0>
{\quad},\eqno(3.15)$$  
one has [29]
$${\mid}\alpha >{\;}{\,}
={\;}{\,}e^{-{\,}{\mid}\alpha{\mid}^{2}/2}{\;}{\,}
\sum^{\infty}_{n=0}{\;}{\,}{{\alpha^{n}}\over{\sqrt{n!}}}{\;}{\,}
{\mid}n>{\quad}.\eqno(3.16)$$  
The coherent state labelled by ${\,}\alpha =0{\,}$ 
is the ground state of the oscillator.  If, for example, 
the system started in a vacuum state, the amplitude 
to find it subsequently in a coherent state 
${\,}{\mid}\alpha>{\,}$ is
$$<0{\mid}\alpha>{\;}{\,}
={\;}{\,}<0{\mid}D(\alpha){\mid}0>{\;}{\,}
={\;}{\,}e^{-{\,}{\mid}\alpha{\mid}^{2}/2}
{\quad},\eqno(3.17)$$  
up to a phase.\par
\smallskip
\indent 
To make complete contact with the amplitude (3.10), 
using coherent-state methods, we note that, 
in terms of the displacement operators $D(\xi){\,}$:
$$<m{\mid}D(\xi){\mid}\alpha>{\;}{\,}
={\;}{\,}{{1}\over{\sqrt{m!}}}{\;}{\,}(\xi +\alpha)^{m}{\;}{\,}
\exp\biggl[-{{1}\over{2}}\Bigl({\mid}\alpha{\mid}^{2}
+{\mid}\xi{\mid}^{2}+2{\,}\xi^{*}\alpha\Bigr)\biggr]
{\quad},\eqno(3.18)$$  
and
$$<m{\mid}D(\xi){\mid}\alpha>{\;}{\,}
={\;}{\,}e^{-{\mid}\alpha{\mid}^{2}/2}{\;}{\,}
\sum^{\infty}_{n=0}{\;}{\,}{{\alpha^{n}}\over{\sqrt{n!}}}{\;}{\,}
<m{\mid}D(\xi){\mid}n>
{\quad}.\eqno(3.19)$$ 
On equating these, one finds that
$$(1+y)^{m}{\;}{\,}e^{-y{\mid}\xi{\mid}^{2}}{\;}{\,}
={\;}{\,}e^{{\mid}\xi{\mid}^{2}/2}{\;}{\,}
\sum^{\infty}_{n=0}{\;}{\,}{\sqrt{{m!}\over{n!}}}{\;}{\;}
\xi^{n-m}{\;}{\,}y^{n}{\;}{\,}<m{\mid}D(\xi){\mid}n>
{\quad}.\eqno(3.20)$$  
But, from the generating function for the associated Laguerre
polynomials [25],
$$(1+y)^{m}{\;}{\,}e^{-yx}{\;}{\,}
={\;}{\,}\sum^{\infty}_{n=0}{\;}{\,}L^{(m-n)}_{n}(x){\;}{\;}
y^{n}{\quad},
{\qquad}{\qquad}{\mid}y{\mid}{\,}<{\,}1
{\quad},\eqno(3.21)$$  
one deduces that the matrix element between initial 
and final states is
$$<m{\mid}D(\xi){\mid}n>{\;}{\,}
={\;}{\,}\biggl({{n!}\over{m!}}\biggr)^{{1}\over{2}}{\;}{\,}
\xi^{m-n}{\;}{\;}e^{-{\mid}\xi{\mid}^{2}/2}{\;}{\;}L^{(m-n)}_{n}
\Bigl({\mid}\xi{\mid}^{2}\Bigr)
{\quad},\eqno(3.22)$$  
which agrees with Eq.(3.10), up to an unimportant phase factor.\par
\medskip
\noindent {\bf 4. Generalised coherent states}
\medskip
\indent 
These amplitudes can also be interpreted in terms of generalised 
coherent states of the harmonic oscillator [28].  Define:
$${\mid}n{\,},\alpha>{\;}{\,}
={\;}{\,}e^{-iE_{n}t}{\;}{\,}D(\alpha(t)){\;}{\mid}n>
{\quad}.\eqno(4.1)$$   
Then, in the Fock representation,
$${\mid}n{\,},\alpha>{\;}{\,}
={\;}{\,}\sum^{\infty}_{m=0}{\;}
<m{\mid}D(\alpha(0)){\mid}n>{\;\,}{\mid}m>{\;}
e^{-iE_{m}t}
{\quad}.\eqno(4.2)$$
For generalised coherent states, the ground state $(n=0)$ 
is a coherent state and not a vacuum state.  Generalised 
coherent states are to the coherent states what 
the Fock states ${\mid}n>$ are to the vacuum state, 
that is, excited coherent states.  In addition, 
denoting by $I$ the identity operator, one finds that [27]:
$$\eqalignno{I&{\;}{\,}
={\;}{\,}{{1}\over{\pi}}{\,}\int{\,}d^{2}\alpha{\;}{\;}
{\mid}n{\,},\alpha>{\;}<n{\,},\alpha{\mid}{\quad},
&(4.3)\cr 
<n{\,},\beta{\mid}n{\,},\alpha>&{\;}{\,}
={\;}{\,}L_{n}\Bigl({\mid}\alpha-\beta{\mid}^{2}\Bigr){\;}{\,} 
\exp\Bigl({\beta^{*}\alpha{\,}-{\,}{{1}\over{2}}
\bigl({\mid}\alpha{\mid}^{2}+{\mid}\beta{\mid}^{2}\bigr)}\Bigr)
{\quad},
&(4.4)\cr
<n{\,},\beta{\mid}\psi>{\;}{\,}&
={\;}{\,}{{e^{-{\mid}\beta{\mid}^{2}/2}}\over{\pi}}{\;}
\int d^{2}\alpha{\;}{\,}L_{n}
\Bigl({\mid}\alpha -\beta{\mid}^{2}\Bigr){\;}
e^{\beta^{*}\alpha}{\;}{\,}e^{-{\mid}\alpha{\mid}^{2}/2}{\;}
<n{\,},\alpha{\mid}\psi>{\quad},
&(4.5)\cr}$$ 
for an arbitrary state ${\,}{\mid}\psi>{\,}$, with the definition:
$${\;}\int d^{2}\alpha{\;}{\,} 
={\;}{\,}\int{\,}d\Bigl[{\rm Re}(\alpha)\Bigr]{\;}{\,} 
d\Bigl[{\rm Im}(\alpha)\Bigr]
{\quad}.\eqno(4.6)$$  
In particular, from Eq.(4.4) with $\beta =0{\,}$, one has
$$<n{\,},0{\mid}n{\,},\alpha>{\;}{\,}
\equiv{\;}{\,}<n{\mid}n{\,},\alpha>{\;}{\,}
={\;}{\,}e^{-{\mid}\alpha{\mid}^{2}/2}{\;\,} 
L_{n}\Bigl({\mid}\alpha{\mid}^{2}\Bigr)
{\;}{\,},\eqno(4.7)$$ 
again giving Eq.(3.6) up to a phase.  The initial state 
should be seen not as a vacuum state, but as a Fock state, 
while the final state should be seen 
as a generalised coherent state.\par
\smallskip
\indent 
As shown by Hollenhorst [30], the amplitudes of Eqs.(3.22) 
have yet a further interpretation:  
they are the matrix elements for a transition 
from state ${\mid}k>$ to state ${\mid}m>$ 
under the influence of a linearised gravitational wave, 
with the force $F(t)$ proportional to the Riemann 
curvature-tensor component $R_{txtx}(t){\,}$:
$$F(t){\;}{\,}
={\;}{\,}\mu{\,}\ell{\,}R_{xtxt}(t){\;}{\,}
={\;}-{\,}{{1}\over{2}}{\,}\mu{\,}\ell{\,} 
\bigl(\partial_{t}\bigr)^{2}h^{TT}_{xx}
{\quad},\eqno(4.8)$$  
where $\ell{\,}$ is the distance between two particles 
along the $x$-axis, each being of mass $(\mu/2){\,}$, 
while ${\,}h^{TT}_{xx}{\,}$ is the transverse-traceless 
gravitational-wave component of the metric [23], 
and $x$ is the change in the separation of the masses.\par 
\smallskip
\indent 
In the context of black-hole evaporation, 
one expects that the role of the force is played 
by the time-dependent background space-time 
-- which approximates a Vaidya space-time 
in the high-frequency limit at late times [4,8,18,19].\par
\smallskip
\indent 
An important point which we should mention is that, 
under the influence of a time-dependent force, 
an initial vacuum state transforms into a coherent state.  
Below, we discuss how, by changing a phase parameter 
of the perturbations appearing in their frequencies 
(parametric amplification), an initial vacuum state 
transforms into a squeezed vacuum state.  
This phase is not an oscillator phase, but a small angle, 
$\delta{\,}$, through which the time $T$ at infinity 
is rotated into the lower complex plane.\par
\medskip
\noindent 
{\bf 5. Squeezed-state formalism}
\medskip
\indent 
In this Section and in the following Sec.6, we shall see how, 
by rotating the asymptotic Lorentzian time $T$ 
into the complex plane, and in the case 
of spherically-symmetric initial matter and gravitational 
fields, one obtains a quantum-mechanical highly-squeezed-state 
interpretation for the final state in black-hole evaporation, 
in the limit of an infinitesimal rotation angle.\par
\smallskip
\indent 
Grishchuk and Sidorov [17] were the first to formulate particle 
creation in strong gravitational fields explicitly in terms 
of squeezed states, although the formalism does appear 
in Parker's original paper on cosmological particle 
production [31].  In [17], it was shown that relic gravitons 
(as well as other perturbations), created from zero-point 
quantum fluctuations as the universe evolves, 
should now be in a strongly squeezed state.  
Squeezing is just the quantum process 
corresponding to parametric amplification.\par
\smallskip
\indent 
Black-hole radiation in the squeezed-state representation 
was first discussed in [17].  The `squeeze parameter' 
$r_{j}$ (see below) was there related to the frequency 
$\omega_{j}$ and the black-hole mass $M$ through
$${\rm tanh}\bigl(r_{j}\bigr){\;}{\,}
={\;}{\,}\exp\Bigl(-{\,}4\pi{\,}M{\,}\omega_{j}\Bigr)
{\quad}.\eqno(5.1)$$  
In this language, the vacuum quantum state in a black-hole 
space-time for each mode is a two-mode squeezed vacuum.  
However, our approach to squeezed states in black-hole 
evaporation is new; arising from a two-surface 
boundary-value problem and Feynman's $+{\,}i\epsilon$ 
prescription [1].  We now give a brief account 
of quantum-mechanical squeezed states.\par
\smallskip
\indent 
A general one-mode squeezed state 
(or squeezed coherent state) is defined [16] as
$${\mid}\alpha{\,},z>{\;}{\,}
={\;}{\,}D(\gamma){\;}S(r{\,},\phi){\,}{\mid}0>{\;}{\,}
={\;}{\,}D(\gamma){\;}S(z){\,}{\mid}0>
{\quad}.\eqno(5.2)$$  
Here, $D(\gamma)$ is the single-mode displacement operator, 
and
$$S(r{\,},\phi){\quad}
\equiv{\quad}S(z){\;}{\,}
={\;}{\,}\exp\biggl({{1}\over{2}}
\Bigl(z{\,}a^{2}-{\,}z^{*}a^{\dagger 2}\Bigr)\biggr)
\eqno(5.3)$$ 
in terms of annihilation and creation operators $a$ 
and $a^{\dagger}{\,}$, respectively, together with the relation
$$z{\;}{\,}
={\;}{\,}r{\,}e^{-2i\phi}
{\quad},\eqno(5.4)$$ 
gives the unitary squeezing operator for ${\mid}\alpha{\,},z>$, 
obeying
$$S^{\dagger}(z){\,}S(z){\;}{\,}
={\;}{\,}S(z){\,}S^{\dagger}(z){\;}{\,}
={\;}{\,}1
{\quad},\eqno(5.5)$$ 
with $\gamma{\,}$ given by
$$\gamma{\;}{\,}
={\;}{\,}\alpha{\,}\cosh{\,}r{\;} 
+{\;}\alpha^{*}{\;}e^{-2i\phi}{\,}\sinh{\,}r
{\quad}.\eqno(5.6)$$  
The state Eq.(5.2) is a Gaussian wave-packet, displaced 
from the origin in position and momentum space.  
While the (real) squeezing parameter 
${\,}r{\;}{\;}(0\leq r<\infty)$ determines 
the magnitude of the squeezing, the squeezing angle
${\,}\phi{\;}{\;}({\,}{\mid}\phi{\mid}<\pi/2{\,})$ 
gives the distribution of the squeezing between 
conjugate variables.  The squeezed vacuum state occurs 
when ${\,}\alpha{\,}={\,}0{\,}$:
$${\mid}z>{\quad}
\equiv{\quad}{\mid}0{\,},z>{\;}{\,}
={\;}{\,}S(z){\,}{\mid}0>
{\quad}.\eqno(5.7)$$  
The limit of high squeezing occurs when $r{\,}{\gg}{\,}1{\,}$, 
where the state ${\,}{\mid}z>$ is highly localised 
in momentum space.\par 
\smallskip
\indent 
Single-mode squeezed operators do not conserve momentum, 
since they describe the creation of particle pairs 
with momentum $k{\,}$.  Two-mode squeezed operators, however, 
describe the creation and annihilation of two particles (waves) 
with equal and opposite momenta.  A two-mode squeeze operator 
has the form [32]
$$S(r,\phi){\;}{\,}
={\;}{\,}\exp\biggl[r\Bigl(e^{-2i\phi}{\;}a_{+}{\;}a_{-}{\,}
-{\,}e^{2i\phi}{\;}a^{\dagger}_{+}{\;}a^{\dagger}_{-}{\,}\Bigr)\biggl]
{\quad},\eqno(5.8)$$  
where ${\,}a_{\pm}$ and ${\,}a^{\dagger}_{\pm}$ are annihilation 
and creation operators for the two modes, respectively.\par
\smallskip
\indent 
Consider two conjugate operators ${\hat p}{\,}$ 
and ${\hat q}{\,}$, with variances $\Delta{\hat p}$ 
and $\Delta{\hat q}{\,}$.  In the squeezed-state formalism, 
one may construct states such that $\Delta{\hat p}{\,}$ 
and $\Delta{\hat q}{\,}$ are equal, taking the minimum value 
possible.  The name `squeezed' refers to the fact 
that the variance of one variable in a conjugate pair 
can go below the minimum allowed by the uncertainty principle 
(the squeezed variable), while the variance of the conjugate 
variable can exceed the minimum value allowed 
(the superfluctuant variable) [25,33,34].  
The superfluctuant variable is amplified by the squeezing 
process, and so becomes possible to observe 
macroscopically, while the subfluctuant variable is squeezed 
and becomes unobservable.  In particle production, 
whether by black holes or in cosmology, the number operator 
is a superfluctuant variable, while the phase is squeezed.\par
\medskip
\noindent 
{\bf 6. Analytic continuation and the large-squeezing limit}
\medskip
\indent 
We shall see here for the black-hole evaporation problem that, 
when one rotates the time-separation $T$ at infinity: 
$T\rta{\mid}T{\mid}\exp(-{\,}i\delta)$ into the complex 
by a very small angle $\delta >0{\,}$, one arrives at a very 
highly-squeezed quantum state.  There is no information-loss 
paradox associated with the relic Hawking radiation, 
as such a state is a pure state.  It is also important 
to state that we do not take the $|T|\to\infty$ limit.  
However, one must understand that the observation time 
at infinity by far exceeds the dynamical collapse time-scale, 
which is of order ${\,}\pi M_{I}{\,}$ [23].  We now repeat Eq.(2.19):
$$\Psi\Bigl[\{A_{j}\};{\,}T\Bigr]{\;}{\,}
={\;}{\,}{\hat N}{\;}e^{-iM_{I}T}{\;}\prod_{j}{\;}\Psi(A_{j};T)
{\quad},\eqno(6.1)$$ 
and then define
$$\eqalign{\Phi\Bigl[\{A_{j}\};{\,}T\Bigr]{\;}{\,}&
={\;}{\,}N{\;}e^{-iM_{I}T}{\,}
\prod_{j}{\;}2i{\,}\sin(k_{j}T){\;}\Psi(A_{j};T)\cr 
&\equiv{\;}{\,}N{\;}e^{-iM_{I}T}{\,}\prod_{j}{\;}
\exp\biggl[{{i}\over{2}}{\,}\bigl(\Delta k_{j}\bigr){\;}k_{j}{\;}
{\mid}A_{j}{\mid}^{2}{\,}\cot(k_{j}T)\biggr]\cr
& ={\;}{\,}N{\;}
\exp\Bigl(i{\,}S^{(2)}_{\rm class}\bigl[\{A_{j}\};{\,}T\bigr]\Bigr){\,}.\cr}
\eqno(6.2)$$ 
\indent 
We further define the functions 
${\,}\phi_{j}\bigl({\,}{\mid}T{\mid}{\,},\delta{\,}\bigr)$ 
and 
${\,}r_{j}\bigl({\,}{\mid}T{\mid}{\,},\delta{\,}\bigr)$ by
$$\phi_{j}\bigl({\,}{\mid}T{\mid}{\,},\delta{\,}\bigr){\;}{\,}
={\;}-{\,}k_{j}{\,}{\mid}T{\mid}{\,}\cos\delta
{\quad},\eqno(6.3)$$
$$\tanh r_{j}\bigl({\,}{\mid}T{\mid}{\,},\delta{\,}\bigr){\;}{\,}
={\;}{\,}\exp\bigl(-{\,}2k_{j}{\,}
{\mid}T{\mid}{\,}\sin\delta{\,}\bigr)
{\quad},\eqno(6.4)$$  
whence
$$\exp(-{\,}2{\,}r_{j}){\;}{\,}
={\;}{\,}\tanh\bigl(k_{j}{\,}
{\mid}T{\mid}{\,}\sin\delta{\,}\bigr)
{\quad}.\eqno(6.5)$$
From Eqs.(6.3-5), one can rewrite Eq.(6.2) in the form
$$\eqalign{\Phi &\Bigl[\{A_{j}\};{\mid}T{\mid}{\,},\delta\Bigr]\cr
&={\;}{\hat N}{\;}e^{-iM_{I}{\mid}T{\mid}\cos\delta}{\;}{\,} 
e^{-M_{I}{\mid}T{\mid}\sin\delta}{\;}\prod_{j}
\exp\Biggl[-{{1}\over{2}}\bigl(\Delta k_{j}\bigr)k_{j}
\biggl({{1+e^{2i\phi_{j}}\tanh r_{j}}
\over{1-e^{2i\phi_{j}}\tanh r_{j}}}\biggr) 
{\mid}A_{j}{\mid}^{2}\Biggr]{\,}.\cr}
\eqno(6.6)$$  
On comparing with Sec.5, we recognise Eq.(6.6) 
as the coordinate-space representation 
of a quantum-mechanical squeezed state [35,36], 
with $r_{j}({\,}{\mid}T{\mid}{\,},\delta{\,})$ 
the squeeze parameter 
and $\phi_{j}({\,}{\mid}T{\mid}{\,},\delta{\,})$ 
the squeeze angle.  The evolution of the squeezed state 
is taken into account by the ${\mid}T{\mid}$-dependence 
in $r_{j}$ and in $\phi_{j}{\,}$, which are in general 
both complicated functions of time.\par 
\ss 
\indent 
We now define
$$\ep_{j}{\;}{\,}
={\;}{\,}k_{j}{\;}{\mid}T{\mid}{\;}\sin\delta{\quad},
{\qquad}{\qquad}{\;}f(k_{j},\ep_{j},{\mid}T{\mid}){\;}{\,} 
={\;}{\,}1{\,}
+{\;}{{\sin^{2}(k_{j}{\,}{\mid}T{\mid})}\over{\sinh^{2}\ep_{j}}}
{\quad}.\eqno(6.7)$$ 
Then
$${\arrowvert}\Phi\bigl[\{A_{j}\};{\mid}T{\mid},\delta\bigr]
{\arrowvert}^{2}{\;}{\,} 
={\;}{\,}{\mid}N{\mid}^{2}{\;}e^{-2M_{I}{\mid}T{\mid}\sin\delta}{\;}
\prod_{j}{\,}\exp\biggl[{{-{\,}\coth\epsilon_{j}}
\over{f(k_{j}{\,},\epsilon_{j}{\,},{\mid}T{\mid})}}{\;}
\bigl(\Delta k_{j}\bigr){\,}k_{j}{\,}
{\mid}A_{j}{\mid}^{2}\biggr]
{\;},\eqno(6.8)$$  
and, from Eqs.(6.5,7):
$$\epsilon_{j}{\quad}
\simeq{\quad}e^{-{\,}2r_{j}}{\quad},
{\qquad}{\qquad}{\quad}\epsilon_{j}{\,}{\ll}{\,}1
{\quad},\eqno(6.9)$$  
corresponding to ${\,}r_{j}{\,}{\gg}{\,}1{\,}$, which is the limit 
of high squeezing.  We discuss the form 
of the normalisation in another paper [9].\par 
\ss 
\indent 
Eq.(6.8) describes a Gaussian non-stationary process 
in which the variance is an oscillatory function of time.  
Rather than dealing with travelling waves, 
one now has standing bosonic waves, where the amplitudes 
for left- and right-moving waves are large and almost equal 
-- this is similar to the inflationary-cosmology 
scenario [17].  One consequence of the high-squeezing 
behaviour is that the variance for the amplitudes 
$\{x_{j}\}$ is large, so that there are large statistical 
deviations of the observable power spectrum 
from its expected value.  This is just a manifestation 
of the Uncertainty Principle.\par 
\ss
\indent
In the squeezed-state formalism, the high-squeezing limit 
${\,}r_{j}{\,}{\gg}{\,}1{\,}$ may be regarded as the classical limit.  
For example, in this sense, in the case of black-hole 
evaporation, the final state of the remnant particle flux 
becomes more classical (more WKB) in the limit 
$\delta\rta 0{\,}$.  In this limit, one can effectively 
consider the final perturbations as being represented 
by a classical probability distribution [17,33,37].  
As in the inflationary scenario in cosmology, 
the perturbations on the spherically-symmetric black-hole 
background space-time, of quantum-mechanical origin, 
cannot be distinguished from classical stochastic 
perturbations, without the need for an environment 
for decoherence.  There is also a correspondence 
between the initial conditions for the perturbations 
in the black-hole and in the cosmological cases.  
In cosmology, the assumption is that, at some early `time' 
just prior to inflation, the modes are in their adiabatic 
ground state.  A similar qualitative statement can be made 
in the black-hole example, provided that the pre-collapse 
initial data were diffuse, slowly-moving 
and spherically symmetric.\par 
\ss 
\indent 
One further consequence follows, 
provided that $\ep_{j}{\,}$ is small (as above).  
Then, one finds for the probability distribution Eq.(6.8) 
that, as ${\,}\delta\rta 0_{+}{\,}$,
$${\arrowvert}\Phi\bigl[\{A_{j}\};{\mid}T{\mid}{\,},\delta\bigr]
{\arrowvert}^{2}{\;}{\quad}
\sim{\quad}{\;}{\mid}N{\mid}^{2}{\;}
\prod_{s\ell mP}{\;}\prod^{\infty}_{n=1}{\;}
\exp\Bigl[-{\,}\bigl(\Delta\omega_{n}\bigr){\;}\omega_{n}{\;} 
{\mid}A_{sn\ell mP}{\mid}^{2}\Bigr]
{\quad},\eqno(6.10)$$ 
where we have used the approximation 
${\,}\sinh\epsilon_{j}{\;}\sim{\;}{\,}\epsilon_{j}{\;}$ for small
${\,}\epsilon_{j}{\;}$, and the identities 
$$\delta(x){\;}{\,}
={\;}{\,}{{1}\over{\pi}}{\,}
\lim_{\epsilon\rightarrow 0}{\,}
{{\epsilon}\over{\bigl({\epsilon}^{2}{\,}+{\,}x^{2}\bigr)}}{\quad},$$ 
and 
$$\delta\bigl[f(x)\bigr]{\;}{\,}
={\;}{\,}\sum_{i}{\;}{\delta(x-x_{i})\over{|f'(x_{i})|}}{\quad},$$ 
where $x_{i}$ are zeros of $f(x)$ and 
${\;}\omega_{n}={\,}n\pi /|T|{\;}$,
${\;}\Delta\omega_{n}=(\omega_{n+1}-\omega_{n}){\;}.$  
We have also used the fact that ${\,}k_{j}\to 0{\,}$ 
and that ${\,}k_{j}{\,}{\mid}A_{j}{\mid}^{2}{\,}\to {\,}0{\,}$ 
as ${\,}k_{j}\to 0{\,}$.  In practice, the product over $n$ 
should be cut off at some large ${\,}n_{max}{\,}$, 
such that ${\,}\omega_{n_{max}}={\,}M_{I}{\,}$.\par 
\ss
\indent 
Further investigation of the derivation of Eq.(6.10) 
indicates that, in the limit of high squeezing, 
the random variable $\phi_{j}$ associated with the final 
state is squeezed to discrete values, independently 
of the quantum numbers $\{s\ell mP\}$ [9].  Note that 
it is only the squeeze phases $\{\phi_{j}\}$ 
of the (standing-wave) perturbations which are fixed 
and correlated in the high-squeezing limit.\par 
\ss 
\indent 
For comparison, in inflationary cosmology, 
the oscillation phases of standing waves have fixed 
values, giving rise to zeros in the power spectrum, 
which are characteristic of the CMBR.  The power spectrum 
of cosmological perturbations in the present universe 
is not a smooth function of frequency.  The standing-wave
pattern, due to squeezing, induces oscillations 
in the power spectrum.  This in turn produces Sakharov 
oscillations [37,38], due to metric and scalar 
perturbations in the distribution of higher-order multipoles 
of the angular correlation function for the temperature 
anisotropies [21,39] in the CMBR, for all perturbations 
at a given time whose wavelength is comparable 
with or greater than the Hubble radius defined for that time.  
That is, the peaks and troughs of the angular power spectrum 
have a close relationship with the maxima and minima 
of the metric power spectrum.  For long wavelengths, 
the power spectrum does become smoother.\par
\medskip
\noindent 
{\bf 7. Conclusion}
\medskip
\indent 
In this Letter, we have illustrated many aspects 
of the quantum boundary-value formulation, 
for linearised bosonic fields (spins $s=0{\,},1{\,},2$) 
propagating in the space-time of an evaporating 
black hole.  When the Lorentzian proper-time 
separation $T$ between the initial and final 
space-like hypersurfaces, as measured at spatial 
infinity, is deformed into the lower complex $T$-plane, 
and when the perturbations are initially weak, 
one obtains a quantum-mechanical squeezed-state formalism.  
The large-squeezing limit is equivalent to the WKB limit, 
corresponding to an infinitesimal angle 
${\,}\delta{\,}{\ll}{\,}1{\,}$ of rotation of $T$ 
into the lower-half complex plane.\par 
\ss
\indent 
Since the final squeezed state is a pure state, 
there is no information-loss paradox as a result 
of the Feynman $+{\,}i\epsilon$ prescription we have adopted.  
Our complex approach is new and differs from Grischchuk's 
original application of squeezed states to black holes.  
However, as in the cosmological scenario, so the bosonic 
perturbations on the black-hole background can be regarded 
as a stochastic collection of standing waves, 
rather than as travelling waves, in the high-squeezing limit.  
This leads to the prediction of peaks in the power spectrum 
of the relic black-hole radiation, analogous to the Sakharov 
oscillations in the CMBR.\par
\goodbreak
\parindent = 1 pt
\medskip
{\bf References}
\medskip
\indent 
[1] R.P.Feynman and A.R.Hibbs, 
{\it Quantum Mechanics and Path Integrals}, 
(McGraw-Hill, New York) (1965).\par 
\indent 
[2] A.N.St.J.Farley and P.D.D'Eath, 
Phys Lett. B {\bf 601} (2004) 184; 
gr-qc/0407086.\par 
\indent 
[3] A.N.St.J.Farley and P.D.D'Eath, 
Phys Lett. B {\bf 613} (2005) 181; 
gr-qc/0510027.\par 
\indent 
[4] A.N.St.J.Farley, 
``Quantum Amplitudes in Black-Hole Evaporation", 
Cambridge Ph.D. dissertation, approved 2002 (unpublished); 
`Squeezed states in black-hole evaporation by analytic continuation' 
(gr-qc/0209113).\par 
\indent 
[5] A.N.St.J.Farley and P.D.D'Eath, 
``Quantum Amplitudes in Black-Hole Evaporation. 
I: Complex Approach", submitted for publication (2006); 
gr-qc/0510028.\par 
\indent 
[6] A.N.St.J.Farley and P.D.D'Eath, 
``Quantum Amplitudes in Black-Hole Evaporation. 
II: Spin-0 Amplitude", submitted for publication (2006); 
gr-qc/0510029.\par 
\indent 
[7] A.N.St.J.Farley and P.D.D'Eath, 
``Bogoliubov Transformations in Black-Hole Evaporation", 
submitted for publication (2006); gr-qc/0510043.\par 
\indent 
[8] A.N.St.J.Farley and P.D.D'Eath, 
``Vaidya Space-Time and Black-Hole Evaporation", 
to appear in Gen. Relativ. Gravit. (2006); gr-qc/0510040.\par 
\indent 
[9] A.N.St.J.Farley and P.D.D'Eath, 
``Quantum Amplitudes in Black-Hole Evaporation: 
Coherent and Squeezed States", to be submitted.\par 
\indent 
[10] A.N.St.J.Farley and P.D.D'Eath, 
``Quantum Amplitudes in Black-Hole Evaporation: Spins 1 and 2'', 
to appear in Ann. Phys. (N.Y.) (2006).\par 
\indent 
[11] Ref.[4], Chapter 6.\par 
\indent 
[12] A.N.St.J.Farley {~} and {~} P.D.D'Eath, 
{~} Class.{~} Quantum {~} Grav. {~} {\bf 22}, {~} 3001  
(2005); gr-qc/0510036.\par 
\indent 
[13] P.R.Garabedian, 
{\it Partial Differential Equations}, 
(Wiley, New York) (1964).\par
\indent 
[14] P.D.D'Eath, 
{~}{\it Supersymmetric {~}Quantum {~}Cosmology}, 
{~}(Cambridge {~}University {~}Press, {~}Cambridge) (1996).\par 
\indent 
[15] R.J.Glauber, 
Phys. Rev. {\bf 131}, 2766 (1963).\par 
\indent 
[16] B.L.Schumacher, 
Phys. Rep. {\bf 135}, 317 (1986).\par 
\indent 
[17] L.P.Grishchuk {~~} and {~~} Y.V.Sidorov, 
{~~~} Phys. {~~~} Rev. {~~~} D{\bf 42}, {~~} 3413 {~~} (1990) {~~} ; 
Class. Quantum Grav. {\bf 6}, L161 (1989).\par 
\indent 
[18] P.C.Vaidya,  
Proc. Indian Acad. Sci. {\bf A33}, 264 (1951).\par
\indent 
[19] R.W.Lindquist, R.A.Schwartz and C.W.Misner, 
Phys. Rev. {\bf 137} 1364 (1965).\par 
\indent 
[20] J.S.Halliwell and S.W.Hawking, 
Phys. Rev D{\bf 31}, 1777 (1985).\par 
\indent 
[21] L.P.Grishchuk, 
Phys. Rev. D{\bf 53}, 6784 (1996).\par 
\indent
[22] J.D.Jackson, {\it Classical Electrodynamics},
(Wiley, New York) (1975).\par
\indent 
[23] C.W.Misner, K.S.Thorne and J.A.Wheeler, 
{\it Gravitation}, (Freeman, San Francisco) (1973).\par 
\indent 
[24] I.S.Gradshteyn and I.M.Ryzhik, 
{\it Tables of Integrals, Series and Products} 
(Academic Press, New York) (1965).\par 
\indent 
[25]  M.Gasperini and M.Giovannini, 
Class. Quantum Grav. {\bf 10}, L133 (1993).\par 
\indent 
[26] J.Schwinger, 
Phys. Rev. {\bf 91}, 728 (1953).\par 
\indent 
[27] S.M.Roy and Virendra Singh, 
Phys. Rev. D{\bf 25}, 3413 (1982).\par 
\indent 
[28] M.V.Satyanarayana, 
Phys. Rev. D{\bf 32}, 400 (1985).\par 
\indent 
[29] K.E.Cahill and R.J.Glauber, 
Phys. Rev. {\bf 177}, 1857, 1882 (1969).\par 
\indent 
[30] J.N.Hollenhorst, 
Phys. Rev. D{\bf 19}, 1669 (1979).\par 
\indent 
[31] L.Parker, 
Phys. Rev. {\bf 183}, 1057 (1969).\par 
\indent 
[32] B.L.Hu, G.Kang and A.L.Matacz, 
Int. J. Mod. Phys. A {\bf 9}, 991 (1994).\par 
\indent 
[33] D.Polarski and A.Starobinskii, 
Class. Quantum Grav. {\bf 13}, 377 (1996).\par 
\indent 
[34] R.Casadio {~~} and {~~} L.Mersini, 
{~~} `Short {~~} distance {~~} signatures {~~} in {~~} cosmology
{~~} : 
Why not in black holes?' (hep-th/0208050).\par 
\indent 
[35] A.L.Matacz, 
Phys. Rev. D{\bf 49}, 788 (1994).\par 
\indent 
[36] C.Kiefer, 
Class. Quantum Grav. {\bf 18}, L151 (2001).\par 
\indent 
[37] A.Albrecht, P.Ferreira, M.Joyce and T.Prokopec, 
Phys. Rev. D{\bf 50}, 4807 (1994).\par 
\indent 
[38] A.Albrecht,
{~~} `Coherence {~} and {~} Sakharov {~} oscillations 
{~} in the {~} microwave {~} sky',
(astro-ph/9612015).\par 
\indent 
[39] T.Prokopec, 
Class. Quantum Grav. {\bf 10}, 2295 (1993).
\end